\documentclass[aps,prb,twocolumn, showpacs]{revtex4}
\usepackage{graphicx}
\usepackage{dcolumn}
\usepackage{amsmath}

\begin{document}

\title{Spin-Hall Effect in A Symmetric Quantum Wells by A Random Rashba Field}
\author{C. P. Moca}
\affiliation{
Department of Physics, University of Oradea, 410087 Oradea, Romania \\
Institute of Physics, Technical University Budapest, Budapest, H-1521,  Hungary
}
\author{D. C. Marinescu}
\affiliation{
Department of Physics and Astronomy, Clemson University, 29634, Clemson
}
\author{S. Filip}
\affiliation{
Department of Physics, University of Oradea, 410087 Oradea, Romania
}
\date{\today}
\begin{abstract}
Changes dopant ion concentrations in the sides of a
symmetric quantum well are known to create a random Rashba-type
spin-orbit coupling. Here we demonstrate that, as a consequence, a
finite size spin-Hall effect is also present. Our numerical
algorithm estimates the result of the Kubo formula for the spin-Hall
conductivity, by using a tight-binding approximation of the
Hamiltonian in the framework of a time-dependent Green's function
formalism, well suited for very large systems.
\end{abstract}
\maketitle

When first discussed almost fifty years ago \cite{Rashba}, the
Rashba-type spin-orbit (SO) coupling was attributed to
the inversion asymmetry in zinc-blende quantum wells grown along the
$[001]$ direction. In this picture, the SO interaction is linear in
the electron momentum $\mathbf{p}$ and is written, in terms of the
Pauli spin operators $\{\sigma_x,\sigma_y,\sigma_z\}$ as:
\begin{equation}
H_{SO} = \alpha (\sigma_xk_y-\sigma_y k_x) \label{eq:rashba}
\end{equation}
The coupling constant $\alpha$ is a constant
proportional to the gradient of the electric potential across 
the well and thus tunable by an electric gate\cite{Nitta}.

Searching for ways and means of manipulating the electron
spin in solid structures by electric fields, within the context of possible spintronics
applications, has rekindled the interest in the Rashba model 
and numerous studies\cite{Sinova, spin_hall} have been dedicated in the the past several years to 
understanding all the implications of this interaction on electron transport 
in reduced dimensionality semiconductor structures. Even though the original
Rashba interaction is intrinsically linked to the quantum well asymmetry, a 
recent argument was made \cite{Sherman1} for the existence of a 
random SO-coupling that appears even in a perfectly
symmetric quantum well on
account of the changes in the dopant ion concentration on the
sides of the well. This result is made possible by the existence of a local
electric field, of random magnitude, perpendicular on the layer, at
each point inside the well. Even though the spatial average of
the random Rashba field that ensues is zero, it has been
demonstrated that it puts a certain imprint on various electronic
properties, such as the spin-relaxation rate which acquires a
minimum value. Moreover, in the presence of a magnetic field, the
effects of the random Rashba fields lead to longer spin relaxation
rates and a non-exponential spin-relaxation.\cite{Sherman2}

These ideas suggest the existence of a finite spin-polarization that
can be maintained also during spin transport. It is quite natural,
therefore, to ask what happens to the spin-Hall conductivity, in
this situation. The present work investigates the effect of the
random Rashba field in symmetric quantum wells on the spin-Hall
conductivity, with the intention of appreciating its robustness
against the natural variations of this type of SO coupling. Within a
numerical algorithm based on the direct integration of the
time-dependent modified Schr\"{o}dinger equation, we establish that indeed, a
minimum spin-Hall effect arises, dependent on the dopant ion
concentration, that is quite resilient under the fluctuations of the
random SO coupling.

\begin{figure}[t]
\centering
\includegraphics[width=3.3in]{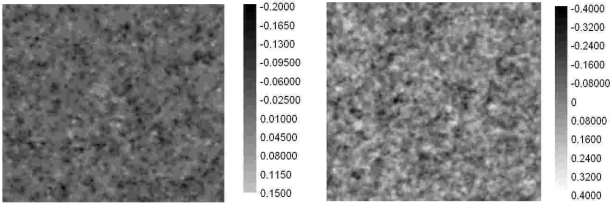}
\centering
\caption{Spatial dependence of the Rashba spin-orbit interaction for different values of $n_u$ and $n_d$, the concentration of carriers on the up and down side of the well, respectively.
Left:  $n_{u}=10^{12} cm ^{-2}$ and $n_{d}=5\times 10^{11} cm ^{-2}$
Right: $n_{d,u}=5\times 10^{12} cm ^{-2}$.
In both figures 
the system size is $400 \times 400 nm$ and $z_0$ 
is fixed to $4nm$.
The energetic units are relative to $t_0$, the hopping integral. }
\label{fig:electric_field_1}
\end{figure}

The physical model of our system consists of a two-dimensional (2D)
quantum well sandwiched between two doped layers of different
concentrations, separated by a distance $z_{0}$. The dopant ions,
assimilated with a $\delta$-type perturbation, are assumed to have
charge $e$ and are localized by a 2D in-plane vector
$\mathbf{r}_{i}= (x_{i},y_{i})$. The total ion concentration
in each layer is $n(\mathbf{r})=\sum_{j}\delta
(\mathbf{r}-\mathbf{r}_{j})$.
In the simplest approximation the Rashba coupling constant is
proportional with the $z$-component of the Coulomb electric field
generated by the impurities:
\begin{equation}
{E}_{z}(\mathbf{r})=\frac{{e}z_{0}}{\epsilon }\sum\limits_{j}\frac{1}{[(\mathbf{r}-\mathbf{r}_{j})^{2}+z_{0}^{2}]^{3/2}},
\end{equation}
where $\epsilon $ is the dielectric constant, and the summation is
performed over all the dopant sites in both layers. Hence,
$\alpha$ depends both on position
$\mathbf{r}$ and on the dopant distribution. In general,  we can
write that $\alpha _{R}(\mathbf{r}) =\alpha _{\mathrm{R}}e{E}_{z}(\mathbf{r})$, where
 $\alpha _{\mathrm{R}}$
is some phenomenological system-dependent constant parameter. An illustration
of this analysis is presented in Fig \ref{fig:electric_field_1} where values of $\alpha _{R}(\mathbf{r})$ are
showed for two different impurity concentrations, for a system size
of $400 \times 400 nm$ and for $z_{0}= 5nm$. 

Since, the position-dependent SO coupling constant $\alpha
_{R}(\mathbf{r})$
 does no longer commute with the momentum operator, the single-particle Hamiltonian that describes the
dynamics of an electron of momentum $\mathbf{p}$ and effective mass
$m^*$ acquires a symmetric form
\begin{equation}
{\tilde H }= \frac{{\mathbf p}^2}{2 m^*} +
\frac{1}{2}\left \{ \alpha_R(\mathbf{r}),
 \sigma _x p_y -\sigma_y p_x \right\},
 \label{eq:hamiltonian}
\end{equation}
where $\left\{A,B \right\}$ represents the anticommutator of  operators $A$ and $B$. The Hamiltonian thus defined in Eq. (\ref{eq:hamiltonian}) contains not just ${ H_{SO}}$, the original Rashba Hamiltonian, Eq.~(\ref{eq:rashba}), written for $\alpha_R(\mathbf{r})$, but also an additional part, $H_{random}$  that reflects
the non-commutativity of momentum and position operators, involved explicitly in $\alpha_R(\mathbf{r})$,
\begin{equation}
{\tilde H_{random}} = \frac{i}{2} \left( {\sigma}_y\frac{\partial\alpha_{R}(\mathbf{r})
}{\partial x}- {\sigma}_x\frac{\partial\alpha_{R}(\mathbf{r})}{
\partial y} \right).
\end{equation}
We note that $\tilde H_{random}$
gives no contribution to other important characteristics of the system, such as the spin-dependent force operator $F_H\sim \alpha_R^2(\mathbf{r})$, which is 
entirely generated by $H_{SO}$. \cite{nikolic}. 

Inspired by Eq.~(\ref{eq:hamiltonian}), we write the Hamiltonian of the total system of electrons directly within the tight binding approximation, generating an expression suitable for numerical calculations:
Thus,
\begin{eqnarray}
H &=&  \sum\limits_{i,\alpha}\varepsilon_{i\alpha} c_{i\alpha}^{\dagger} c_{i\alpha}
-\sum\limits_{<i,j>,\alpha} t_{0} c_{i\alpha}^{\dagger}c_{j\alpha}\\
&+& \sum\limits_{i, \delta_x, \delta_y} V_R\left ( i\right) \left[\left( c_{i\uparrow}^{\dagger}c_{i+\delta_x \downarrow}-
c_{i\downarrow}^{\dagger}c_{i+\delta_x \uparrow}\right)\right . \nonumber \\
& &-i\left. \left( c_{i\uparrow}^{\dagger}c_{i+\delta_y \downarrow}+
c_{i\downarrow}^{\dagger}c_{i+\delta_y \uparrow}\right) \right].
\nonumber \label{eq:tot-ham}
\end{eqnarray}
Here $c_{i\alpha}^{\dagger}$ is the creation operator at site index 
$i$ with spin $\alpha$,  $\delta_x$ and $\delta_y$ are unit vectors
along the $x$ and $y$ directions. 
An immediate consequence of the randomness of the Rashba spin-orbit strength is the site-dependence of the
parameters that describe the model. First, an on-site energy, $\epsilon_{i\alpha}$, that depends on the gradient components of
the electric field appears, similar to the case of disorder. The difference, however, is that in
this case the disorder is spatially correlated. Then, the Rashba coupling $V_R(i)= \alpha_R(i)/a_0 $, written for a lattice constant $a_0$, depends on the values of the electric fields at neighbor sites.  The hopping coupling $t_0$ has the usual expression $t_0 = 1/2m^* a_0^2$. In the following estimates, we fix its value at $t_0= 12 meV$ 
which  corresponds to a lattice constant  $a_0\simeq 2nm$. (Throughout this work, $\hbar =1$,  and free boundary conditions are used.) At the same time 
the charge concentration fixes the Fermi energy $\epsilon_F$ through the relation: $n = 2m\epsilon_F/2\pi \hbar^2= 10^{12}cm^{-2}$. 
A numerical estimate of the spin-Hall conductivity in a random
environment, for large system sizes, can be obtained by using the Kubo formalism in
the framework proposed by Tanaka \cite{Tanaka}. This method can be,
in principle, used for the calculation of any expectation value of
any combinations of Green's functions and quantum operators, such as
the density of states, or conductivity\cite{Iitaka1, Tanaka}.
Besides, it is well suited for large systems where traditional
methods such as direct diagonalization fails due to memory problems.
Within this algorithm, for a given field distribution, the on-site energies, hopping probabilities and the
Rashba field are computed first at each site. Then, the spin-Hall conductivity is
calculated for a given configuration.

The main steps involved in performing the numerical calculations are
outlined below. The computation starts from solving the time-dependent
modified Schr\" odinger equation, with a single-frequency source term:
\begin{equation}
i\frac{ d\, \left | \tilde{j}, t \right >}{d\, t} = H \left | \tilde{j}, t \right > + \left|j\right>\theta(t)
\exp^{-i(E+i\eta)t}\;, \label{eq:schr}
\end{equation}
where $\eta$ is a finite small value and $\theta$ is the step function.

To determine the time evolution of the state ket $\left
|\tilde{j};t\right
>$, a direct numerical integration of the modified Schr\" odinger equation is
performed, using the "leap-frog" algorithm\cite{Iitaka}. This is a
second order, symmetrized differencing scheme, accurate up to
$(H\Delta t)^2$. In this approximation, Eq.~(\ref{eq:schr}) becomes:
\begin{eqnarray}
\left|\tilde{j};t+\Delta t\right>&=&-2i\Delta t H\left|\tilde{j};t\right> 
+\left|\tilde{j};t-\Delta t\right>\\ \nonumber
&-& 2i\Delta t \left|j\right> e^{-i(E + i\eta)t}\theta (t)\;.
\end{eqnarray}
The time step $\Delta t$ is determined by $\Delta t \simeq \beta/|E| $, where $|E|$ is the absolute value of the
energy and $\beta$ is a parameter whose value is less than 1 in order for the solution to be stable.

An analytic solution of Eq.~(\ref{eq:schr}), with the initial
condition $\left | \tilde{j}, t=0 \right >=0$ is written as
\begin{eqnarray}
\left | \tilde{j}, t \right > & = & -i\int_{0}^{t}dt'e^{-iH(t-t')}\left | {j}\right >
e^{-i(E+i\eta)t'}\nonumber\\
& = &
\frac{1}{E+i\eta-H}\left[e^{-i(E+i\eta)t}-e^{-iHt)}\right]\left | {j}\right >\;,\label{eq:green1}
\end{eqnarray}
where one recognizes the Green's function operator $(E+i\eta-H)^{-1}$ as the prefactor in the final expression.
For sufficiently large time, the Green's function operating on the ket $\left | {j}\right >$
can be obtained with the relative accuracy $\delta = e^{-\eta T}$ by inverting Eq.~(\ref{eq:green1})
as:
\begin{equation}
G(E+i\eta)\left | {j}\right >=\lim_{T\rightarrow\infty}\left |\tilde {j}, T\right >e^{ i(E+i\eta)T}.
\end{equation}
Consequently, the matrix element between any two states $\left<i\right|$ and $\left|j\right>$ is obtained as:
\begin{equation}
\left<i\right|G(E+i\eta)\left|j\right> = \lim_{T\rightarrow \infty}\left< i \right|\left.\tilde {j}, T\right >\;.
\end{equation}

These results can be easily generalized to the case when a
product including several Green's
functions and operators is involved by choosing a new
initial state, such as $\left|j'\right> = AG(E+i\eta)\left|j\right>$
in Eq.~(\ref{eq:schr}) and repeating the same procedure. In a similar fashion the method can be generalized to
the case when many different energy values are considered. In this case
one solves Eq.~(\ref{eq:schr})
simultaneously for a source term with multiple frequencies,
 $\left|j\right>\left(\sum_l e^{-i(E_l+i\eta)t}\right)\theta(t)$.

The efficiency of the algorithm is increased if instead of a local orbital basis set it
 a randomized version of this
basis  is selected. This can be described described by a ket $\left| \phi \right > = \sum_{n=1}^{N}
\left | n \right > \exp(-i \, \phi_n)$, where $\left | n \right >$
are the localized orbitals and $\phi_n$ are random numbers in
the $\left[0, 2\pi \right]$ interval. Then, the average of a given operator $A$ is calculated as
\begin{equation}
\left < \phi\right| A\left | \phi \right >\simeq \sum_n \left < n
\right |  A\left| n \right > + O\left(\frac{1}{\sqrt{N}}\right)\;,
\end{equation}
within the statistical errors of $1/\sqrt{N}$.
It was showed\cite{Iitaka2} that this choice of basis is the
best one for reducing the numerical errors.

\begin{figure}[h]
\centering
\includegraphics[width=3.3in]{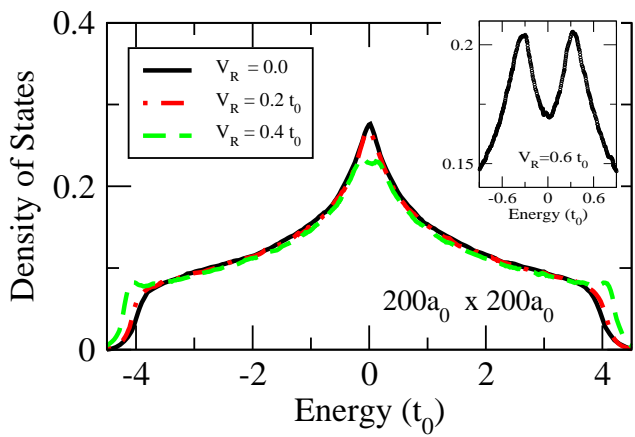}
\caption{The density of states of a clean system with constant Rashba coupling $V_R$ is  calculated for a square lattice of size $ 200\times 200 a_0 $. The energy is expressed in hopping-integral units, $ t_0 $.}
\label{fig:density_of_states}
\end{figure}

As a first application of the present method we calculate the
density of states (DOS),
\begin{equation}
\varrho\left( \omega\right) = -\frac{1}{\pi} \Im m\left[  Tr
G\left(\omega+i\eta \right)  \right] \label{eq:dos}
\end{equation}
 in a system size $200a_0\times 200 a_0$ for a constant $V_R$. The results, computed for three different values of $V_R$ are presented in Fig. \ref{fig:density_of_states}. 
In the presence of a constant Rashba field, the band structure
suffers two important modifications. First, the bandwidth increases
proportional to $V_R$ as a consequence of the
renormalization of the hopping probability by the Rashba interaction. 
Second, the van-Hove singularity in the density
of states that occurs at zero energy when $V_R =0$ splits when
$V_R$ is turned on. This corresponds to the splitting and shifting of
the original spin-degenerate band into the two chiral sub-bands, characteristic to the Rashba model . Analytically
it can be shown\cite{winkler} that the 
splitting is proportional with $V_R^2 $. This behavior can be observed
in the inset of Fig.\ref{fig:density_of_states} where a large
$V_R=0.6t_0$ was considered.

The spin-Hall conductivity can be calculated with the Kubo
formula
\begin{eqnarray} \sigma _{sH}=\frac{1}{2}\, {\rm Tr} \int
\frac{d\varepsilon }{2\pi } \left( -\frac{\partial f(\varepsilon
)}{\partial \varepsilon }\right) \left< {j}_x^z \left[
{G}_R(\varepsilon )-{G}_A(\varepsilon )\right] \right. \nonumber \\
\left. \times \, {v}_y\, {G}_A(\varepsilon ) -{j}_x^z\,
{G}_R(\varepsilon )\, {v}_y \left[{G}_R(\varepsilon
)-{G}_A(\varepsilon )\right] \right>\;.
\label{eq:kubo_spin_hall_conductivity}
\end{eqnarray}
This expression considers contributions only from states at the Fermi surface,
as indicated by the presence of the sharply peaked derivative of the
distribution function. Such an approximation is justified by
previous studies of the spin-Hall effect, which demonstrated that
that states below the Fermi surface does not contribute to the
spin-Hall conductivity and that spin-Hall conductivity comes
entirely from quasiparticle states at the Fermi
level\cite{dimitrova}.

In Eq.~(\ref{eq:kubo_spin_hall_conductivity}), the velocity operator
is defined by the commutator: $i\, v_y = \left[ y, H \right
]$, whereas the spin current, along the $\hat{x}$ direction
 given in terms of the anticommutator between the velocity
operator and the Pauli matrix $\sigma_z$: $j_x^{z} =\left\{
\sigma_z, v_x \right\}/4$\cite{Niu}. $G_{R/A}(\epsilon)$ represents
the retarded/advanced Green's function.

\begin{figure}[h]
\centering
\includegraphics[width=3.3in]{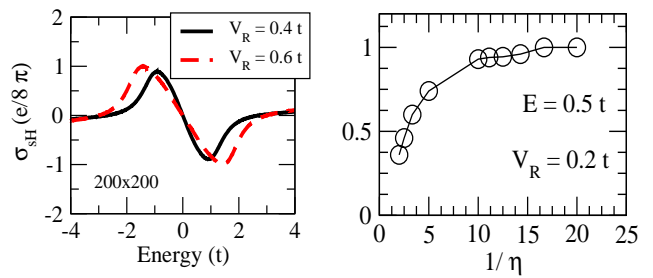}
\caption{Left: The spin Hall conductivity for a 
clean system of size $ 200\times 200 a_0 $,
with a constant $V_R$. The results are obtained 
by averaging over 200 initial wave functions. 
Right: $\eta$ dependence of the spin-Hall
conductance for a fixed energy $E=0.5 t$ and a Rashba 
coupling $V_R = 0.2 t$. The error bars are smaller 
than the symbols size.}
\label{fig:spin_hall_clean_system}
\end{figure}

For a constant Rashba field, the  algorithm outlined above generates results, 
which after being averaged over 200 different impurity distributions, are
presented in Fig. \ref{fig:spin_hall_clean_system}.  
For any value of
$V_R$, the universally predicted value of $\sigma_{sH}=e/8\pi$ is
reached, but only for some energy interval inside the band. 
Usually
the spin-hall conductivity is smaller than $e/8\pi$. It decreases as
the band edges are approached and vanishes beyond them. This
behavior is preserved for system sizes up to $500a_0\times 500a_0$,
indicating that the system remains in the ballistic regime
regardless of its size for as long as no disorder is included. 
In the right inset of Fig. \ref{fig:spin_hall_clean_system} we present 
the $\eta$ dependence of the spin-Hall conductivity. For values 
smaller than 0.1 already the convergence to the approapriate value is reached. 
Increasing $\eta$ above 0.1 deviations are already consistent even if a 
larger integration time is used. Throughout of our calculations we have used
$\eta =0.1$.
Similar results for the spin-Hall conductance 
were obtained using a Landauer-B\" uttiker formalism\cite{nikolic}.

\begin{figure}[h]
\centering
\includegraphics[width=3.3in]{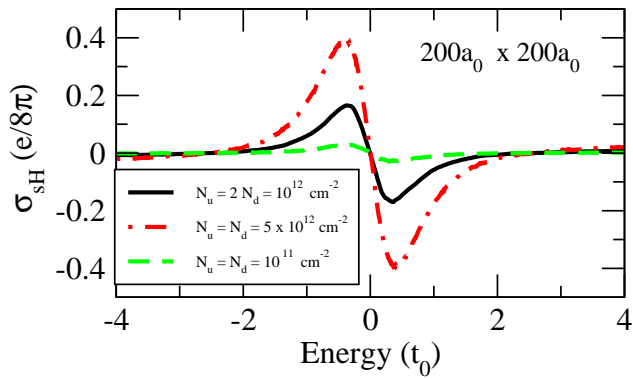}
\caption{The spin Hall conductivity when both up-down doping
layers are present. For each Rashba field distribution
the spin-Hall conductivity was averaged over 200
initial wave functions at each energy. The figure presents
averages over 100 random impurity configurations.}
\label{fig:spin_hall_random_rashba}
\end{figure}

Calculated values of the spin-Hall conductivity in the presence of a
random Rashba field, averaged over $100$ impurity configurations, are showed in Fig.
\ref{fig:spin_hall_random_rashba} for different dopant 
concentration of the sides of the well, $u$ (up) and $d$ (down) respectively. 
Here, the randomness of the
electric field created by the dopants generate a Rashba coupling at
each lattice site, even in the absence of an external Rashba field.
A finite $\sigma_{sH}$ is still obtained. The parameters used for
our analysis are experimentally reachable\cite{Nitta}. The solid and
dashed-dotted lines in Fig. \ref{fig:spin_hall_random_rashba}
correspond to the spatial distribution of the Rashba fields
presented in Fig. \ref{fig:electric_field_1}. 

When  a symmetric distribution of the
impurity ions is considered on the sides of the well, a vanishingly
small Rashba field ensues when averaged over the entire sample, 
as shown in \ref{fig:electric_field_1}. The corresponding spin-Hall conductivity, 
however, which is determined by the configuration average of the spatial 
variation of the coupling constant $\alpha$ increases as expected with 
the impurity concentration and seems to be favored by perfectly symmetric distributions. 
When the numbers of impurities is different on the
sides of the quantum well, the average Rashba field is finite, but the 
spin-Hall conductivity decreases.
A possible explanation of this result can be given in 
terms of the spatial correlation of the spin-orbit interaction 
over the spin precession length\cite{Sherman1}  that are strong 
enough to assure the existence of a finite spin-Hall effect even 
in the case of a null average spin-orbit field.  

In conclusion we have studied the spin-Hall effect in a perfectly
symmetric quantum well when fluctuations of the Rashba spin-orbit
interaction field are considered as arising due to the impurities on
the sides of the well. Even in the extreme limit, when an equal
number of the impurities generate a Rashba field averaged to a very
small value, a finite spin-hall conductivity seem to exist in the
system in the range of $5\div 10 \%$ of the original universal value
$e/8\pi$.

{\it Acknowledgments -} One of the authors (CPM) acknowledge support from the Hungarian Grants OTKA
Nos. NF061726 and T046303, and by the Romanian Grant CNCSIS 2007/1/780. Discussions
with E. Ya. Sherman are acknowledged.

\end{document}